\documentstyle[12pt,amssym]{article}
\def\R{\mbox{$\Bbb R$}}
\def\i{\mbox{\rm i}}
\def\case#1#2{{\textstyle{#1\over #2}}}

\title{
\hfill{\normalsize ULB/229/CQ/00/2}\\
\vspace{1cm}
PT-symmetric sextic potentials}

\author{B.\ Bagchi $^{a,}$\thanks{E-mail: bbagchi@cucc.ernet.in}\ , F.\ Cannata
$^{b,}$\thanks{E-mail: Francesco.Cannata@bo.infn.it}\ , C.\ Quesne
$^{c,}$\thanks{Directeur de recherches FNRS; E-mail: cquesne@ulb.ac.be} \\
{\small \sl $^a$ Department of Applied Mathematics, University of Calcutta,} \\
{\small \sl 92 Acharya Prafulla Chandra Road, Calcutta 700 009, India}\\
{\small \sl $^b$ Dipartimento di Fisica and INFN, Via Irnerio 46, 40126 Bologna,
Italy}\\
{\small \sl $^c$ Physique Nucl\'eaire Th\'eorique et Physique
Math\'ematique,  Universit\'e Libre de Bruxelles,} \\ {\small \sl Campus de la
Plaine CP229, Boulevard~du Triomphe, B-1050 Brussels, Belgium}}
\date{ }
\begin{document}
\baselineskip=22pt plus 1pt minus 1pt
\maketitle

\begin{abstract}
The family of complex PT-symmetric sextic potentials is studied to show that for
various cases the system is essentially quasi-solvable and possesses real,
discrete
energy eigenvalues. For a particular choice of parameters, we find that under
supersymmetric transformations the underlying potential picks up a
reflectionless
part.
\end{abstract}

\vspace{0.5cm}






\newpage
%
%
\section{Introduction}

Searching for non-Hermitian PT-symmetric Hamiltonians has acquired much
interest in recent times (see e.g.~\cite{bender,andrianov,znojil,bagchi} and
references quoted therein). For one thing, a rather large subclass of such
Hamiltonians has been found to possess real energy eigenvalues. For
another, in at
least some cases, it is seen that a complex shift of the coordinate $x \in
(-\infty,
\infty)$ does not affect the overall normalizability of the wave functions,
while at
the same time retaining the real character of the energy spectrum.\par
%
%
The purpose of this letter is twofold:

\noindent
(i) We examine the general problem of a complex sextic potential from the
point of
view of determining exactly a finite number of eigenvalues and eigenfunctions. A
suitable ansatz scheme leads us to find discrete real energy levels under quite
general conditions.

\noindent
(ii) We point out that some of our results can also be arrived at by
performing a
complex shift of the coordinate on the reduced sextic potential consisting of
even-power terms only. However, our results cover a much greater ground. In
particular, we find it possible to generate an additional complex reflectionless
term in the sextic potential by employing supersymmetric transformations.\par
%
%
\section{Complex sextic potentials and their solutions}

To get started, let us consider the following general representation of a
sixth-degree potential
\begin{equation}
  V(x) = \sum_{i=1}^6 c_i x^i,  \label{eq:V}
\end{equation}
satisfying the Schr\"odinger equation (in units $\hbar = m =1$)
\begin{equation}
  \left[- \frac{1}{2} \frac{d^2}{dx^2} + V(x)\right] \psi(x) = E \psi(x),
\end{equation}
where for $V(x)$ to be PT symmetric, $c_1$, $c_3$, $c_5 \in \i\R$, but $c_2$,
$c_4$, $c_6 \in \R$.\par
%
%
We make the ansatz that the wave function is of the form
\begin{equation}
  \psi(x) = f(x) \exp\left(- \sum_{j=1}^4 b_j x^j\right),
\end{equation}
where $f(x)$ is some polynomial function of $x$, which, for complex
potentials, is
typically of the type $\sum_{m=0}^n \alpha_m (\i x)^m$. For the real analogue
of~(\ref{eq:V}) consisting of even power terms only, $f(x)$ is known to have
a given parity.\par
%
%
We focus on the following choices of $f$:
\[\begin{array}{ll}
     \mbox{(a)} & f(x) = 1, \\[0.2cm]
     \mbox{(b)} & f(x) = x + a_0, \\[0.2cm]
     \mbox{(c)} & f(x) = x^2 + a_1 x + a_0,
  \end{array}\]
but can generalize to higher degrees as well. For complex potentials, $a_0$ is
imaginary in (b), whereas $a_1$ is imaginary, but $a_0$ is real in (c).\par
%
%
Without going into the details of calculations, which are quite
straightforward, let
us summarize our results.\par
%
%
\subsection{\boldmath The $f(x)=1$ case}

Here the potential parameters are found to be related to the $b$'s as
\begin{eqnarray}
  c_1 & = & - 3 b_3 + 2 b_1 b_2, \qquad c_2 = - 6 b_4 + 3 b_1 b_3 + 2
b_2^2, \qquad
         c_3 = 4 b_1 b_4 + 6 b_2 b_3, \nonumber \\
  c_4 & = & 8 b_2 b_4 + \case{9}{2} b_3^2, \qquad c_5 = 12 b_3 b_4, \qquad
         c_6 = 8 b_4^2.
\end{eqnarray}
\par
%
%
Without loss of generality, we can choose $c_6 = \frac{1}{2}$ to fix the leading
coefficient of $V(x)$. It gives $b_4 = \pm \frac{1}{4}$. We take the
positive sign to
ensure normalizibility of the wave function, which reads
\begin{equation}
  \psi(x) = \exp\left(- b_1 x - b_2 x^2 - b_3 x^3 - \case{1}{4} x^4\right).
  \label{eq:wf-constant}
\end{equation}
The associated energy level is given by
\begin{equation}
  E = b_2 - \case{1}{2} b_1^2.
\end{equation}
\par
%
%
Now $b_1$ and $b_3$ imaginary make $c_1$, $c_3$, $c_5$ imaginary too, so we
have a complex PT-symmetric potential. Note that the energy eigenvalue in such a
case is real, and the corresponding wave function is PT-symmetric.\par
%
%
\subsection{\boldmath The $f(x) = x + a_0$ case}

The wave function is of the form
\begin{equation}
  \psi(x) = (x + a_0) \exp\left(- b_1 x - b_2 x^2 - b_3 x^3 - \case{1}{4}
x^4\right)
\end{equation}
for
\begin{eqnarray}
  c_1 & = & - 6 b_3 + 2 b_1 b_2 + a_0, \qquad c_2 = - \case{5}{2} + 3 b_1 b_3
         + 2 b_2^2, \qquad c_3 = b_1 + 6 b_2 b_3, \nonumber \\
  c_4 & = & 2 b_2 + \case{9}{2} b_3^2, \qquad c_5 = 3 b_3, \qquad
         c_6 = \case{1}{2}.
\end{eqnarray}
There is also a condition on $a_0$,
\begin{equation}
  a_0^3 - 3 b_3 a_0^2 + 2 b_2 a_0 - b_1 = 0.  \label{eq:condition}
\end{equation}
The energy is given by
\begin{equation}
  E = - \case{1}{2} b_1^2 + 3 b_2 - 3 a_0 b_3 + a_0^2.
\end{equation}
\par
%
%
Let us discuss some important special cases of this scheme.\par
%
%
\subsubsection{$b_1 = b_3 =0$}

The condition~(\ref{eq:condition}) reduces to
\begin{equation}
  a_0 \left(a_0^2 + 2 b_2\right) = 0.  \label{eq:condition-bis}
\end{equation}
\par
%
%
(i) If $a_0 = 0$, then there is no imaginary term in the potential, and this
corresponds to the $n=0$, negative-parity result of ref.~\cite{turbiner}.
The energy
eigenvalue is given by $E = 3 b_2$ and shows a single level.\par
%
%
(ii) The other solution of~(\ref{eq:condition-bis}), namely $a_0^2 = -
2b_2$ can be
studied according to whether $a_0^2 > 0$ or $a_0^2 < 0$.\par
%
%
If $a_0^2 > 0$, then a linear term is present in $V(x)$ with $c_1 = \pm \sqrt{2
|b_2|}$, $c_2 = 2 b_2^2 - \frac{5}{2}$, and $c_4 = 2 b_2$. Of course
$c_3 = c_5 = 0$. The energy is $E = b_2 < 0$. Thus we have two different real
potentials with the same energy eigenvalue. The linear term breaks PT invariance
of the potential and the wave function as well. So, in this respect, $a_0$
real can be
viewed as an explicit symmetry breaking parameter.\par
%
%
On the other hand, if $a_0^2 < 0$, we get two different complex potentials,
corresponding to $c_1 = \pm \i \sqrt{2b_2}$, with the same real energy
eigenvalue:
\begin{eqnarray}
  V(x) & = & \case{1}{2} x^6 + 2 b_2 x^4 + \left(2 b_2^2 -
\case{5}{2}\right) x^2
         \pm \i \sqrt{2b_2}\, x, \label{eq:V-linear} \\
  E & = & b_2 > 0, \\
  \psi(x) & = & \left(x \pm \i \sqrt{2b_2}\right) \exp\left(- b_2 x^2 -
\case{1}{4}
         x^4\right), \label{eq:wf-linear}
\end{eqnarray}
The potential~(\ref{eq:V-linear}) is PT symmetric, while the wave
function~(\ref{eq:wf-linear}) is odd under PT symmetry.\par
%
%
\subsubsection{$b_1 = 0$, $b_3 \ne 0$}

The solution for $a_0=0$ turns out to give the same conclusions as previously
obtained. The second solution
\begin{equation}
  a_0 = \frac{1}{2} \left(3 b_3 \pm \sqrt{9 b_3^2 - 8 b_2}\right) \label{eq:a_0}
\end{equation}
yields two possibilities according as $b_3 \in \R$ or $b_3 \in \i \R$. If
$b_3 \in
\R$, we have $b_3^2 \ge \frac{8}{9} b_2$, implying two different real potentials
with the same energy eigenvalue, except for the equality sign
in~(\ref{eq:a_0}).\par
%
%
If however $b_3 \in \i \R$, then $a_0$ must be imaginary with $b_3^2 = - |b_3|^2
\le \frac{8}{9} b_2$. Here too we have two possibilities of obtaining two
different
complex potentials with the same real energy eigenvalue, except for the equality
sign in~(\ref{eq:a_0}).\par
%
%
\subsection{\boldmath The $f(x) = x^2 + a_1 x + a_0$ case}

The complete set of solutions leading to more than one energy level
corresponds to
\begin{equation}
  a_1 = 2 b_3, \qquad a_0 = \frac{1}{2} \left(2 b_2 - b_3^2 \pm
\sqrt{\left(2 b_2
  - 3 b_3^2\right)^2 + 2}\right),  \label{eq:a_1-a_0}
\end{equation}
and gives
\begin{eqnarray}
  V(x) & = & \case{1}{2} x^6 + 3 b_3 x^5 + \left(2 b_2 + \case{9}{2}
b_3^2\right) x^4
          + 2 b_3 \left(4 b_2 - b_3^2\right) x^3 + \left[2 \left(b_2^2 + 3
b_2 b_3^2
          - 3 b_3^4\right) - \case{7}{2}\right] x^2 \label{eq:V-quadratic}
\nonumber \\
  && \mbox{} + b_3 \left(4 b_2^2 - 4 b_2 b_3^2 - 7\right) x,
          \label{V-quadratic} \\
  E_{\pm} & = & - 2 b_3^2 \left(b_2 - b_3^2\right)^2 + 3 b_2 - b_3^2 \pm
          \sqrt{\left(2 b_2 - 3 b_3^2\right)^2 + 2}, \\
  \psi_{\pm} & = &\left[x^2 + 2 b_3 x + \case{1}{2}\left(2 b_2 - b_3^2 \mp
          \sqrt{\left(2 b_2 - 3 b_3^2\right)^2 + 2}\right)\right] \nonumber \\
  && \mbox{} \times \exp\left[- 2 b_3 \left(b_2 - b_3^2\right) x - b_2 x^2
- b_3 x^3
          - \case{1}{4} x^4\right].  \label{eq:wf-quadratic}
\end{eqnarray}
\par
%
%
The results (\ref{eq:a_1-a_0})--(\ref{eq:wf-quadratic}) are valid both for
real and
imaginary $b$'s. In the latter case, PT symmetry is good for the potential
and the
wave function, whereas in the former one has a symmetry breaking. Note that in
both cases $E_+ > E_-$.\par
%
%
Concerning the case where $b_3$ is imaginary, we have a complex PT-symmetric
two-parameter family of potentials with two distinct real energy levels. This is
truly a non-trivial result and puts the spirit of quasi-solvability in the
complex
domain. Indeed, for the particular case of $b_3 = 0$ and $b_2 = \gamma/2$, we
recover the $n=1$, positive-parity results of the one-dimensional even-power
potential $V(x) = \frac{1}{2} x^6 + \gamma x^4 + \case{1}{2} \left(\gamma^2 +
\mu\right) x^2$, where $\mu = -3 - 4n - 2r$ and $r$ is associated with the
$(-1)^r$
parity of $n+1$ levels.\par
%
%
The converse also works. The results (\ref{eq:a_1-a_0})--(\ref{eq:wf-quadratic})
can be derived from the even-power sextic potential of ref.~\cite{turbiner} by a
translation $x \to x + b_3$. If $b_3$ is imaginary, then this translation
amounts to
a complex shift, whose viability has already been pointed out in
refs.~\cite{andrianov,znojil}.\par
%
%
\section{A supersymmetric viewpoint}

In order to enlarge the class of PT-invariant potentials, SUSY methods have been
used thoroughly~\cite{andrianov,bagchi}. Here, we also outline the procedure to
generate superpartners of the potentials considered in the previous
section, which
share their PT properties. The procedure is based on the construction of a
superpotential $W(x) = - \psi'(x)/\psi(x)$, and therefore is sound in so
far as the
logarithmic derivative of $\psi(x)$ is well behaved on the real axis.\par
%
%
One can start from the wave function~(\ref{eq:wf-constant}), the parameters
$b_1$
and $b_3$ being assumed imaginary. Then the superpartner
\begin{equation}
  \tilde{V}(x) = V(x) - \left(\frac{\psi''(x)}{\psi(x)} - \frac{\psi^{\prime
 2}(x)}{\psi^2(x)} \label{eq:V-partner}
  \right)
\end{equation}
becomes again a sextic potential, and there is no real enlargement.\par
%
%
More interesting is the case where one starts from the wave
function~(\ref{eq:wf-linear}) and constructs the partner
of~(\ref{eq:V-linear}). By
substituting (\ref{eq:wf-linear}) into~(\ref{eq:V-partner}), we find
$\tilde{V}$ to
be
\begin{equation}
  \tilde{V}(x) = V(x) + \left(\frac{1}{x \pm \i \sqrt{2b_2}}\right)^2 + 2
b_2 + 3 x^2,
  \label{eq:V-transparent}
\end{equation}
where the piece in parentheses is clearly reflectionless.  The latter is indeed
reminiscent of the ``transparent'' complex potential obtained in
ref.~\cite{andrianov}, which is invariant under PT and gives a trivial
$S$-matrix.
Moreover this piece has an associated zero-energy bound state, given by
$\Psi_0(x)
= C \left(x \pm \i \sqrt{2b_2}\right)^{-1}$, where $C$ is a
constant.\footnote{This
transparent potential may also be viewed as the particular case $l=1$ of the
generic potential $\frac{1}{2} l (l+1) \left(x \pm \i
\sqrt{2b_2}\right)^{-2}$ that
is typical of a centrifugal barrier in a radial context (but with no
singularity) and
which has a zero-energy bound state $\Psi_0(x) = C \left(x \pm \i
\sqrt{2b_2}\right)^{-l}$.}\par
%
%
The form~(\ref{eq:V-transparent}) also generalizes the result obtained for the
harmonic oscillator in ref.~\cite{znojil} to the sextic case. The fact that
$b_2 \ne
0$ makes the potential rather appealing in that if $b_2$ were vanishing, there
would be a singularity on the real axis and the potential would be rendered
ill defined. Here, because of a shift of the singularity,
(\ref{eq:V-transparent})
remains well posed in the complex plane cut from $x = \mp \i \sqrt{2b_2}$ to $x
= \mp \i \infty$, respectively.\par
%
%
Finally, one can consider the wave function~(\ref{eq:wf-quadratic}) and
construct
the partner of~(\ref{eq:V-quadratic}). Again one can see that
$\tilde{V}(x)$ is not
trivial: for $b_3$ imaginary, $\psi'(x)/\psi(x)$ is well behaved on the
real axis.\par
%
%
All results we have obtained can be generalized to the case where one shifts the
variable $x$ by a translation $b$, with $b$ real. In such a case, the
potentials and
wave functions are reparametrized correspondingly. One may worry however about
PT properties since one can generate subleading powers in $x$ from a given
power.
However, one should realize that now the parity operation can be defined with
respect to a mirror placed at $x = -b$, so that $x + b = x - (-b)$ goes to
$- x - 2b -
(-b) = - (x + b)$. When this translation $b$ is performed, the reflectionless
potential contained in~(\ref{eq:V-transparent}) becomes precisely that
considered
in ref.~\cite{andrianov}.\par
%
%
\section{Conclusion}

To conclude, we have solved a complex PT-symmetric sextic potential in its most
general form within a suitable ansatz scheme for the wave functions and
shown how
the associated energy levels turn out to be real. We have also demonstrated
using
SUSY the possibility of generating a complex reflectionless part in the
potential.
Although we have restricted our discussion up to the quadratic order in the
coefficient of the exponential representing the wave function, it is
obvious that we
can build up, in an identical way, higher-order states.\par
%
%
\section*{Acknowledgement}

One of us (B.B.) thanks Professor C.\ Quesne for warm hospitality at ULB,
Brussels.
He also thanks FNRS and University of Calcutta for providing financial
support.\par
%
%

\end{document}